\documentclass{article}
\usepackage{spconf,amsmath,graphicx, xcolor, subcaption, amsfonts, multirow}

\usepackage{enumitem}
\setlist{nosep, leftmargin=14pt}

\usepackage{mwe} %

\title{Quantifying Hippocampal Shape Asymmetry in Alzheimer's Disease using Optimal Shape Correspondences}

\name{Shen Zhu$^{\star \dagger}$ , Ifrah Zawar$^{\ddagger}$, Jaideep Kapur$^{\ddagger}$, P. Thomas Fletcher$^{\star \dagger}$}
\address{$^{\star}$  Department of Computer Science, University of Virginia, Charlottesville, VA, USA\\
    $^{\dagger}$Department of Electrical and Computer Engineering, University of Virginia, Charlottesville, VA, USA\\
    $^{\ddagger}$Department of Neurology, University of Virginia School of Medicine, Charlottesville, VA, USA
    }
\begin{document}
\maketitle
\begin{abstract}
Hippocampal atrophy in Alzheimer's disease (AD) is asymmetric and spatially inhomogeneous.
While extensive work has been done on volume and shape analysis of atrophy of the hippocampus in AD, less attention has been given to hippocampal asymmetry specifically. Previous studies of hippocampal asymmetry are limited to global volume or shape measures, which don't localize shape asymmetry at the point level. 
In this paper, we propose to quantify localized shape asymmetry by optimizing point correspondences between left and right hippocampi within a subject, while simultaneously favoring a compact statistical shape model of the entire sample.
To account for related variables that have an impact on AD and healthy subject differences, we build linear models with other confounding factors.
Our results on the OASIS3 dataset demonstrate that compared to volumetric information, shape asymmetry reveals fine-grained, localized differences that inform us about the hippocampal regions of most significant shape asymmetry in AD patients.
\end{abstract}
\section{Introduction}
\label{sec:intro}

Alzheimer's disease (AD) is the most common type of neurodegenerative disorder and is characterized by cognitive decline \cite{magalingam2018current, zawar2023does}. 
The hippocampus is one of the most common sites of neurodegeneration in AD and is often atrophied \cite{sarica2018mri}. 
Hippocampal atrophy in AD is often asymmetric. The asymmetry of the hippocampus 
in pathological AD brains is a well-known phenomenon that may have diagnostic, management, and prognostic implications and can dictate cognitive trajectory \cite{stefanits2012asymmetry}. Thus, understanding the hippocampal asymmetry can be clinically informative. While the asymmetry of hippocampus in histopathology has long been studied in AD \cite{stefanits2012asymmetry}, anatomical asymmetry on structural neuroimaging has received recent attention. Fox \textit{et al.} \cite{fox1996presymptomatic} documented asymmetrical atrophy of the brain for people with AD. In a more recent study \cite{sarica2018mri}, the authors revealed that the asymmetry index of hippocampus increases with the progression of AD.

While the analysis of asymmetry in brain structure volumes provides information about atrophy in AD, volume measurements are a crude summary of the complex neuroanatomy. Volumetric analysis on conventional magnetic resonance imaging (MRI) can provide little information about specific regions within the hippocampus and how their involvement in the disease process may impact cognitive trajectories. In contrast, statistical shape analysis captures finer localized neuroanatomic variations that can have more meaningful clinical and prognostic implications.
Thus, by exploring asymmetry of the brain with shape analysis, we could potentially better characterize and visualize the morphological asymmetric changes in the structures of interest in people with AD.

There has been very limited research on characterizing brain shape asymmetry in AD. Wachinger et al.~\cite{wachingerWholebrainAnalysisReveals2016} used \textit{BrainPrint}~\cite{wachingerBrainPrintDiscriminativeCharacterization2015} to analyze the shape asymmetry of several brain structures in AD. This method uses surface-based Laplace-Beltrami eigenvalues and eigenfunctions to define shape descriptors called ShapeDNA~\cite{reuter_laplacebeltrami_2009}. The authors demonstrated that using ShapeDNA to characterize morphological asymmetry outperforms using volumetric asymmetry on several structures.
However, using Laplace-Beltrami eigenvalues and eigenfunctions to characterize shape has its drawbacks. Wachinger \textit{et al.} \cite{wachingerWholebrainAnalysisReveals2016} tried to localize areas of the hippocampal asymmetry that show the strongest association with AD. They utilized level-set analysis, in which the level set curves from the first eigenfunction wrap around the hippocampus. They computed a mixed-effects model on the absolute difference of level-set curve lengths for the left and right hippocampus.
The level-set analysis relies on the assumption that the corresponding eigenfunction for each subject are in correspondence, which does not always hold because of the eigenvector switching problem \cite{zhang2010spectral}. Additionally, level-set curve length is an aggregated measure that lacks point-wise correspondences. The visualization of shape differences is limited to highlighting level set curves with significant asymmetry in their lengths. Thus, there is not a point-wise and intuitive statistical analysis and visualization of the asymmetry.

Therefore, to better understand the fine-grained morphological asymmetry differences between healthy controls and AD patients, 
we propose to use a point distribution model (PDM) \cite{cates2007shape} as the representation for hippocampal shape. In addition, we devise a method for point-wise quantification and statistical analysis of the hippocampal asymmetry, which enables localization of the areas with significant shape asymmetry differences between healthy controls and people with AD. Our shape model is based on the particle correspondence optimization method of ShapeWorks~\cite{cates2007shape}, which we augment to optimize interhemispheric correspondences of left and right hippocampi within subjects at the same time as optimizing shape correspondences across subjects. We devise a point-wise shape asymmetry measure and statistical analysis that we show on OASIS3~\cite{lamontagne2019oasis} data. Our findings highlight statistically significant fine-grained shape asymmetry in AD.

\section{Method}
\subsection{Preliminaries}

ShapeWorks \cite{cates2007shape} is a software tool that constructs compact particle-based shape models on a sample of anatomical structures. It iteratively optimizes the positions of particles on the surfaces of each subject's anatomical structure. The optimization achieves two goals: a) evenly distributing the particles on each surface so as to fully capture the surface geometry, and b) maintaining geometric correspondences of particles across the set of surfaces.

These two goals are achieved by minimizing a joint cost function $Q$. For a sample of $N$ 3D shapes, $S=\{z_1, ..., z_N\}$, each shape $z_i$ has $M$ particles that characterize the surface, denoted as $z_i = [z_i^1, ..., z_i^M] \in \mathbb{R}^{3M}$, with each particle living in $\mathbb{R}^3$. There are two types of random variables in this particle-based representation: a \textit{shape space} random variable $\mathbf{Z} \in \mathbb{R}^{3M}$, and a \textit{particle position} random variable $\mathbf{X}_i \in \mathbb{R}^{3}$ that encodes the distribution of particles on the $i$-th shape.

By minimizing the joint cost function:
\begin{equation}
\label{equa:entropy_cost}
    Q=\alpha H(\mathbf{Z}) - \sum_{i=1}^N H(\mathbf{X}_i),
\end{equation}
where $H$ is an entropy estimation of the distribution, and $\alpha$ is the relative weighting, ShapeWorks aims to create a statistical shape model with compact \textit{shape space} distribution by minimizing the first term of \eqref{equa:entropy_cost}, which enforces particle correspondence among shapes, and good surface sampling for each shape by maximizing the second term of \eqref{equa:entropy_cost}.

\subsection{Proposed Approach}
\label{sec:proposed_approach}

The first step is to build the PDM of hippocampus for $N$ subjects.
In order to have good correspondences between left and right hippocampi and across the population,
we flip the segmentation of right hippocampus for each subject along the sagittal plane, and use the left hippocampus segmentation and flipped right hippocampus segmentation as input to ShapeWorks \cite{cates2017shapeworks}. We note that \eqref{equa:entropy_cost} is invariant to the flipping of all input, thus the choice of flipping either the left or right side will yield the same optimal correspondences.

We then rigidly align the input shapes to a chosen reference shape, and set the number of points $M$ on each hippocampus to be 512, which is heuristically selected to ensure good surface sampling. The output shape models from ShapeWorks are registered to the mean shape model through generalized Procrustes analysis iteratively until the mean shape model converges. To characterize the shape asymmetry for subject $i$, we subtract each point $l_i^m (m=1, ..., 512) \in \mathbb{R}^3$ of the left hippocampus by the \textit{corresponding point} $r_i^m$ of the flipped right hippocampus ($l_i$ and $r_i$ are the $z_i \in \mathbb{R}^{3\times512}$ computed by ShapeWorks for left and right hippocampi), and get the difference vector $\mathbf{d}_i^m$ from the right to left side:
\begin{equation}
    \mathbf{d}_i^m = l_i^m - r_i^m.
\end{equation}

The normal component of the difference vector encodes the inward or outward movement of the surface, which is related to hippocampal atrophy. Whereas the tangential component accounts for the particle sliding along the surface, which does not describe localized atrophy. Thus, we would like to use a reference shape for calculating the normal directions.
Here, we use the linear mean shape, $E_i$, of the left PDM $l_i$ and the PDM of flipped right hippocampus $r_i$ as the reference for the $i$th subject, i.e.,
\begin{equation}
    E_i = (l_i + r_i) / 2.
\end{equation}

The outward-pointing normal direction of $E_i$ at point $m$ for subject $i$, denoted $\mathbf{n}_i^m$, is calculated from a dense mesh estimated by ShapeWorks. 
For subject $i$, the difference at point $m$ is defined as:
\begin{equation}
    y_i^m = \mathbf{d}_i^m \cdot \mathbf{n}_i^m,
\end{equation}
where $y_i^m$ is the directional change from the left side to the right side. To characterize undirectional changes, we only need to take the absolute value $|y_i^m|$, and the directional asymmetry for each subject $i$ is characterized by $y_i = [y_i^1, ..., y_i^M] \in \mathbb{R}^M$.

Finally, to account for other factors that might affect asymmetry, we used a linear model to include other factors as covariates. Initially, we chose sex, age, handedness, and estimated intracranial volume (eTIV) as covariates. However, handedness was later removed in the final linear model after we found out that it was not considered as statistically significant. 
We denote the covariates for subject $i$ (age at the scan, sex, eTIV, and diagnosis) as $X_{ij} (j=1, 2, 3, 4)$ respectively.
The linear model for the point asymmetry $y_i^m (i=1, ..., N)$ at point $m$ as dependent variable is:
\begin{equation}
\label{equa:linear_model}
    y_i^m = \beta_0^m + \beta_1^m X_{i1} + \beta_2^m X_{i2} + \beta_3^m X_{i3} + \beta_4^m X_{i4}.
\end{equation}
We extract the statistics and $p$-values of covariate diagnosis $X_{i4}$ in each linear model for our analysis, and $p$-values are false discovery rate (FDR) thresholded by $q=0.05$.

\section{Results}

\subsection{Data}

The data we used in this study is from OASIS3 database 
\cite{lamontagne2019oasis}. Since we aimed to compare neuroimaging findings of people with AD with healthy controls, we determined the subject type using the clinical diagnosis in the database. Eventually, we included 208 AD patients and 717 healthy participants with normal cognition who met the inclusion criteria. The segmentation results and statistics are provided in OASIS3.

\subsection{Volumetric Asymmetry Analysis}

\begin{figure}
    \centering
    \includegraphics[width=1\linewidth]{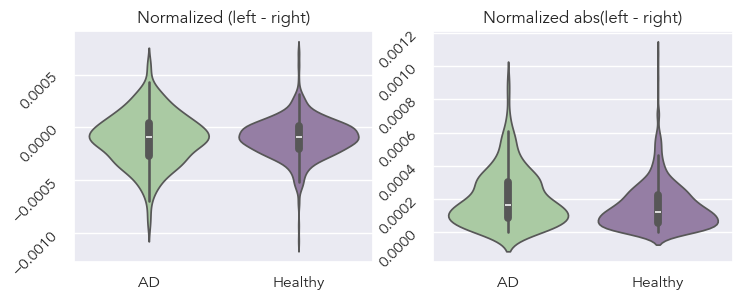}
    \caption{Normalized directional volume asymmetry and undirectional volume asymmetry}
    \label{fig:volumetric_diff}
\end{figure}

We first looked into the hippocampus volumetric asymmetry characteristics between AD and healthy control groups. The volumes were normalized by dividing by eTIV. We analyzed both normalized directional left-to-right asymmetry, and normalized undirectional asymmetry, as shown in Figure \ref{fig:volumetric_diff}.

The distributions of the undirectional asymmetry measures are highly skewed. Thus normality cannot be assumed, which we confirmed with a Shapiro-Wilk normality test, and a $t$ test is not appropriate. Therefore, we chose to use a non-parametric Mann-Whitney U test. From Table \ref{tab:mannwhitneyutest}, we can see that we were able to reject the null hypothesis for normalized undirectional difference but not the directional difference. This shows that using volumetric information alone is not enough to characterize the directional asymmetry between two groups. Thus, it is necessary for us to investigate the directional asymmetry using shape analysis.

To further validate our conclusion, we also built a multivariate linear model with the same covariates as in \eqref{equa:linear_model}, i.e., sex, age and eTIV, and we used normalized directional and undirectional asymmetry as dependent variable. Even after accounting for these variables, we found significant difference in undirectional asymmetry but not in directional asymmetry, with $p$-values being $1.80\times10^{-4}$ and $0.33$ respectively.

\begin{table}
    \centering
    \begin{tabular}{|c|cc|cc|}
    \hline
    \multirow{2}{*}{}     & \multicolumn{2}{c|}{\textbf{\begin{tabular}[c]{@{}c@{}}Normalized\\ (left - right)\end{tabular}}} & \multicolumn{2}{c|}{\textbf{\begin{tabular}[c]{@{}c@{}}Normalized\\ abs(left - right)\end{tabular}}} \\ \cline{2-5} 
                          & \multicolumn{1}{c|}{\textbf{AD}}                        & \textbf{Healthy}                        & \multicolumn{1}{c|}{\textbf{AD}}                          & \textbf{Healthy}                         \\ \hline
    \textbf{Mean} ($\times 10^{-4}$)& \multicolumn{1}{c|}{-1.04}                              & -0.96                                   & \multicolumn{1}{c|}{2.06}                                 & 1.57                                     \\ \hline
    \textbf{SD} ($\times 10^{-4}$)& \multicolumn{1}{c|}{2.44}                               & 1.84                                    & \multicolumn{1}{c|}{1.66}                                 & 1.36                                     \\ \hline
    \textbf{U-statistics} & \multicolumn{2}{c|}{73712.0}                                                                      & \multicolumn{2}{c|}{61725.0}                                                                         \\ \hline
    \textbf{$P$-values}     & \multicolumn{2}{c|}{0.80}                                                                         & \multicolumn{2}{c|}{\textbf{0.00015}}                                                                \\ \hline
    \end{tabular}
    \caption{Summary of asymmetry and Mann-Whitney U tests}
    \label{tab:mannwhitneyutest}
\end{table}

\subsection{Shape Asymmetry Analysis}

\begin{figure}
    \centering
    \includegraphics[width=0.9\linewidth]{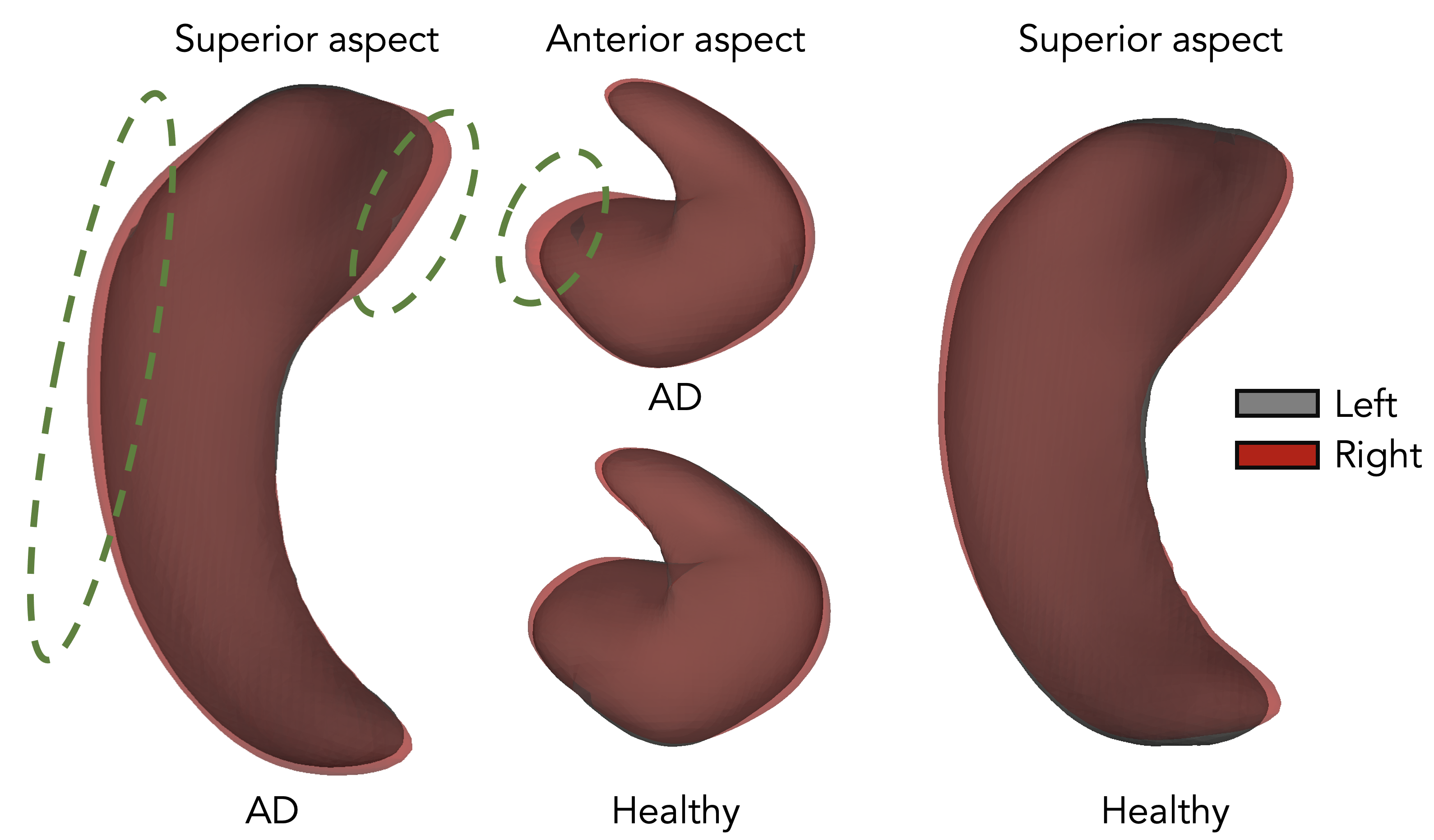}
    \caption{Superimposed left and right mean hippocampus for AD and healthy subjects}
    \label{fig:mean_left_and_right}
\end{figure}

As an overview of the shape asymmetry, we first qualitatively assessed the asymmetry by superimposing the left and flipped right mean hippocampus for both AD and healthy controls group (Figure \ref{fig:mean_left_and_right}). The left side was found to be generally more atrophied at the head and lateral side.

We then examined the mean asymmetry for both groups. 
Each subject has asymmetry $y_i = [y_i^1, ..., y_i^{512}] \in \mathbb{R}^{512}$. We took the linear mean asymmetry for both AD and healthy as the group asymmetry.
To quantitatively assess the asymmetry between the two groups, we first performed principal component analysis (PCA) and then a Hotelling $T^2$ test. The PCA reduces dimensionality and regularizes the covariance estimates needed for the $T^2$ test. The number of components to retain was determined by Horn's parallel analysis, resulting in 30 components that explain $92.90\%$ of the variance. 
The test statistic was $\hat{T}^2 = 168.37$, with corresponding $p$-value being numerically zero, which showed strong statistically significant difference in shape asymmetry between the two groups.

\begin{figure}
    \centering
    \includegraphics[width=1\linewidth]{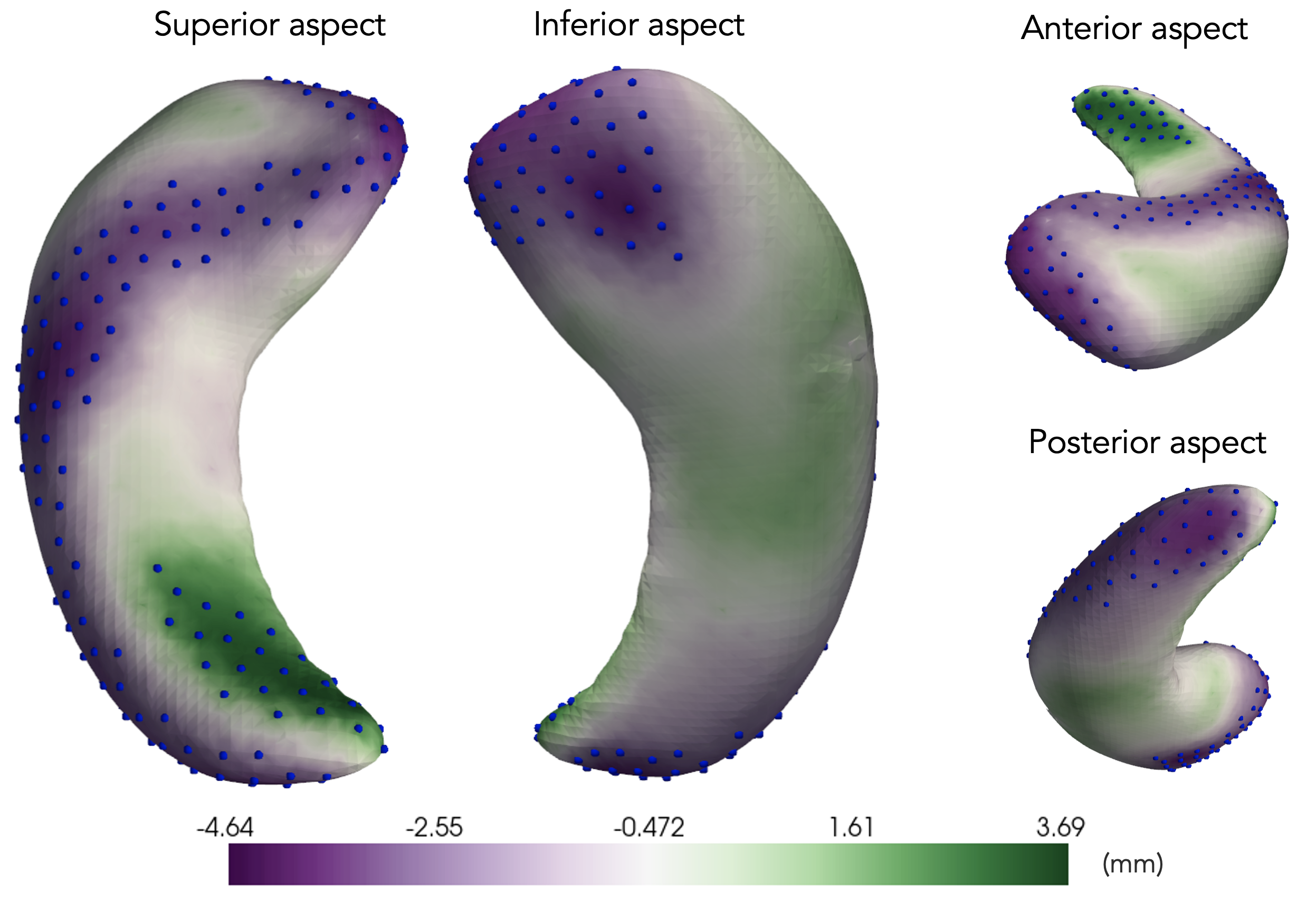}
    \caption{Linear model statistics of diagnosis and statistically significant areas after FDR correction(blue dots)}
    \label{fig:fdr_corrected}
\end{figure}
We note that shape asymmetry highlights the {\it directionality} of hippocampal asymmetry, which is not possible using volumetric analysis alone.
With the Hotelling test analyzing the overall shape asymmetry, we can further analyize the point-wise shape asymmetry components that significantly contribute to differentiating the two groups. In order to do so, we built a linear model as shown in \eqref{equa:linear_model} for each point $m$, with sex, age, eTIV, and diagnosis as covariates, as discussed in Section~\ref{sec:proposed_approach}. We extracted the $p$-values and statistics for covariate diagnosis $X_{i4}$ in each linear model. Finally, $p$-values were corrected using FDR with threshold $q=0.05$. We found out that, out of 512 tests, we were able to reject 177 null hypothesis that correspond to 177 points out of 512 that are considered as statistically significant in differentiating AD and healthy groups. The areas that are covered by these 177 points are also viewed as the most statistically significant areas of directional shape asymmetry, and they characterize the shape asymmetry changes in the course of healthy to AD disease progression.

\section{Discussion}

Compared to the level-set method, our method was able to visualize the fine-grained shape asymmetry difference in a combined manner, instead of on each eigenfunction. This method of visualization is more intuitive and could be extended to other cortical and subcortical structures in the brain.

In volumetric asymmetry analysis, we found undirectional asymmetry to be significantly different but not directional asymmetry when we compared people with AD to healthy controls.
Sarica \textit{et al.} \cite{sarica2018mri} had similar findings for undirectional asymmetry.
In the directional shape asymmetry analysis, we found significant asymmetry in AD compared to controls, and hippocampal head and lateral portions were more atrophied on the left side in AD compared to the right side.

In conclusion, we proposed a method to characterize the hippocampal shape asymmetry in people with AD, which is highly sensitive and enables intuitive and combined visualization. This can have significant clinical, diagnostic, management and prognostic implications.

\section{Acknowledgments}
\label{sec:acknowledgments}
This research received funding support from Alzheimer's Association (Grant Number: AACSFD-22-974008, PI: Dr. Zawar). This work was partially supported by NSF Smart and Connected Health grant 2205417.

\section{Compliance with Ethical Standards}
\label{sec:compliance}

This research study was conducted retrospectively using human subject data made available in open access in OASIS3~\cite{lamontagne2019oasis}. Ethical approval was not required as confirmed by the license attached with the open access data.

\bibliographystyle{IEEEbib}
\bibliography{refs_first_author}

\end{document}